\documentclass[aps,pre,showpacs,twocolumn]{revtex4-1}
\usepackage{graphicx}

\begin{document}

\title[]{Quantum phase transition of the transverse-field quantum Ising model on scale-free networks}
\author{Hangmo \surname{Yi}}
\email{hyi@ssu.ac.kr}
\affiliation{Department of Physics, Soongsil University, Seoul 156-743, Korea}
\affiliation{Institute for Integrative Basic Sciences, Soongsil University, Seoul 156-743, Korea}

\begin{abstract}
I investigate the quantum phase transition of the transverse-field quantum Ising model in which nearest neighbors are defined according to the connectivity of scale-free networks. Using a continuous-time quantum Monte Carlo simulation method and the finite-size scaling analysis, I identify the quantum critical point and study its scaling characteristics. For the degree exponent $\lambda=6$, I obtain results that are consistent with the mean-field theory. For $\lambda=4.5$ and 4, however, the results suggest that the quantum critical point belongs to a non-mean-field universality class. Further simulations indicate that the quantum critical point remains mean-field-like if $\lambda>5$, but it continuously deviates from the mean-field theory as $\lambda$ becomes smaller.
\end{abstract}


\maketitle

\section{Introduction}

Recently, complex networks have drawn much attention among the physics community because they can describe a huge variety of both natural and man-made systems, such as the World Wide Web,\cite{albert1999a,huberman1999a} the Internet,\cite{faloutsos1999a} human social networks,\cite{newman2001a,render1998a,liljeros2001a} power grids,\cite{watts1998a} ecological networks,\cite{montoya2002a,garlaschelli2003a} metabolic networks,\cite{jeong2000a} neural networks,\cite{watts1998a,amaral2000a} and protein interactions\cite{jeong2001a}.
In a physical model based on such a network, individual elements interact with each other if they are directly connected by a link.
The equilibrium properties of such a system are strongly affected by its topological structure.
As opposed to a regular lattice with nearest neighbor interactions, a complex network has random long-range connections that enhance long-range correlation, clustering, and small-worldness.
While the ``volume'' of the system is proportional to the number of nodes $N$, the largest distance grows no faster than $\log N$, hence the effective dimension of a complex network becomes infinitely large.\cite{albert2002a}
In many cases, these effects are manifested in critical phenomena, where the increase of long-range connections drive the critical point toward the mean-field universality class.\cite{scalettar1991a,dorogovtsev2008a}

The degree $k$ of a node is defined as the number of links it is connected to.
The way the degrees are distributed is an important property that characterizes a complex network.
Interestingly, in a large proportion of the widely observed complex networks, the degree distribution follows a power law, $P(k)\propto k^{-\lambda}$.\cite{albert2002a}
These special kinds of complex networks are called scale-free networks because no particular length scale can be found in the degree distribution.
The direct consequence of the power-law distribution is that there are always a small but significant fraction of nodes with large degrees.
These ``hub'' nodes play an important role of reducing the average distance between the nodes, and they greatly influence the phase transitions.
Critical behavior of the Ising model on scale-free networks has recently been a popular subject of research.
It is now very well-known that the universality class of the ferromagnetic-paramagnetic phase transition depends on the degree exponent $\lambda$.\cite{dorogovtsev2002a,leone2002a,yi2008a,yi2010a}
For $\lambda>5$, the phase transition is of the mean-field type.
If $3<\lambda<5$, however, the critical point does not belong to the mean-field universality class, and its critical exponents change depending on the value of $\lambda$.
For $\lambda<3$, the critical temperature $T_c$ becomes infinitely large, and no phase transition is possible at finite temperatures.
This ``tunability'' of the universality class is one of the reasons that the scale-free networks are often studied in the context of critical phenomena.

Recently, there have been efforts to extend the Ising model on scale-free networks to include quantum effects.\cite{yi2008a,yi2010a}
As a magnetic field $\Delta$ is introduced in the direction perpendicular to the Ising spin direction, it gives rise to quantum fluctuations which tend to weaken the spin-spin correlation of the system.
This effect is best manifested by the fact that $T_c$ decreases monotonically with growing $\Delta$.
If $T_c$ vanishes at a finite transverse field $\Delta_c$, this is a quantum critical point, at which rich and interesting phenomena of the quantum phase transition are observed.\cite{sachdev1999booka}
In general, the quantum critical point belongs to a different universality class from that of the classical critical point.
In some special cases, however, both kinds of critical points may belong to the same universality class.
Most notably, if the classical critical point belongs to the mean-field universality class, so does the quantum critical point.\cite{baek2011a}

While there are so many examples of classical dynamic systems on scale-free networks, the study of quantum systems is mostly theoretical at the moment. As the technology of reducing the system sizes advances so rapidly, however, the quantum effect may become practically relevant in the near future.
Another motivation for the study of quantum model on complex networks is related to the recent development in the quantum computing, for which the effect of imperfections in the control of local magnetic field at each node is quite an important issue.
For example, a recent study of a quantum mechanical model on a completely random network\cite{song2001a} has shown that the quantum computational errors grow very fast with the number of quantum bits.
How the topology of the underlying network affects this property is a very intriguing question in itself, and the study of quantum model on complex networks may contribute to the understanding of such effect.

The quantum critical point of the transverse-field quantum Ising model has been studied in various structures including the globally coupled network,\cite{botet1982a,botet1983a} Watts-Strogatz small-world networks,\cite{baek2011a} the Bethe lattice,\cite{nagaj2008a,krzakala2008a} and fractal lattices.\cite{yi2015a}
One of the important properties of the quantum critical point is the dynamic critical exponent $z$.
This quantity determines how the temporal dimension should scale compared to the spatial dimensions, in order the keep the action invariant under renormalization.
The effective total dimension of the quantum critical system in $d$ spatial dimensions thus becomes $d+z$.
In particular, it is known that $z$ is equal to one for the transverse-field Ising model on any integer-dimensional regular lattices.\cite{sachdev1999booka}
This theorem also applies to any quantum critical point in the mean-field universality class, because the upper critical dimension in this case is an integer.\cite{baek2011a}
Fractal lattice models are one of the known examples for which $z$ is not equal to one.\cite{yi2015a}

In recent studies on the critical behavior of the transverse-field quantum Ising model at {\em finite} $T_c$, it was confirmed that the presence of the transverse magnetic field does not affect the universality class of the critical point.\cite{yi2008a,yi2010a}
Although the {\em quantum} critical point of the model has not been studied so far, there are a few predictions we can make.
First, the quantum critical point for $\lambda>5$ is expected to be in the mean-field universality class, since the dimension of the classical critical point is already above the upper critical dimension.
The properties of the quantum critical point for $\lambda<5$ is, however, far from obvious.
Yet we can think of two possible scenarios.
Either the addition of the temporal dimension would drive the critical point to the mean-field universality class, or it will still belong to a non-mean-field universality class.
In the latter case, the universality class of the quantum critical point is most likely different from that of the classical critical point.

In this paper, I will study the behavior of the quantum critical point of the transverse-field Ising model on scale-free networks, using a quantum Monte Carlo simulation program that I developed.
First of all, I use the Suzuki-Trotter decomposition method\cite{suzuki1976a} to write the action as an integral in the imaginary time.
This may be thought of as mapping the quantum model into a classical model with an additional dimension.
The size of the imaginary time dimension is inversely proportional to $T$, and therefore becomes infinitely large at the quantum critical point.
In order to deal with the infinite size of the system, I will apply the finite-size scaling method.

Among many available quantum Monte Carlo simulation techniques,\cite{rieger1999a,yi2003a,sandvik2003a} my program is based on a continuous-time method developed by Rieger and Kawashima.\cite{rieger1999a}
The main procedure may be summarized as follows.
Before the simulation begins, the world line of length $\beta\equiv 1/(k_BT)$ for each spin is broken into segments with a random length and spin direction.
The position of the starting point of each segment, which may take any arbitrary real number between 0 and $\beta$, is stored in an array.
At the start of each Monte Carlo step, the segments are further divided by new cuts that are inserted at random positions according to a Poisson process.
After clusters are formed by neighboring equal-spin segments according to a probability determined by their temporal overlap, a random spin direction is assigned to each cluster.
This is most efficiently accomplished using the Swendesen-Wang cluster algorithm.\cite{swendsen1987a,wang1990a}
After removing redundant cuts between segments with the same spin direction, the procedure goes back to the beginning of the next Monte Carlo step.
This cycle is repeated until the desired accuracy is achieved.

The outline of this paper is as follows.
In Sec.~\ref{sec:model}, the transverse-field quantum Ising model is introduced and explained along with the details of how I construct the scale-free networks.
In Sec.~\ref{sec:results}, the results of the simulations are presented and analyzed for a few values of $\lambda$, both above and below five.
Then I conclude with summary and discussions in Sec.~\ref{sec:summary}.

\section{Model}
\label{sec:model}

The Hamiltonian of the transverse-field quantum Ising model is given by
\begin{equation}
  H = -J \sum_{\left<ij\right>} \sigma^z_i \sigma^z_j + \Delta \sum_i \sigma^x_i  
\end{equation}
where $\sigma^x_i$ and $\sigma^z_i$ are Pauli matrices representing the $x$ and $z$ components of the spin at the $i$th node.
I will consider only the ferromagnetic case ($J>0$), and use the unit in which $J=1$ and $k_B=1$ for simplicity.
The first summation runs over all connected pairs of nodes in the scale-free network under consideration.
Note that the second term does not commute with the first, hence causes quantum fluctuations to the energy eigenstates of the first Hamiltonian term alone.

The ensemble of the scale-free networks used in this research is defined by three parameters: the degree exponent $\lambda$, the total number of nodes $N$, and the average degree $k_\mathrm{av}$.
More specifically, the degree distribution probability is given by
\begin{equation}
  P(k) = \left\{ \begin{array}{ll}
      0, & $if $ k<k_\mathrm{min} \\
      P_0, & $if $ k=k_\mathrm{min} \\
      ck^{-\lambda},\ & $if $ k>k_\mathrm{min},
  \end{array} \right.
\end{equation}
where $k_\mathrm{min}$, $P_0$, and $c$ are determined by the conditions
\begin{equation}
  \sum_{k=1}^\infty P(k) = 1 \quad \mathrm{and}\quad \sum_{k=1}^\infty kP(k) = k_\mathrm{av}.
\end{equation}
Note that $k_\mathrm{min}$ plays the role of the lower cutoff in the degree.
In order to ensure that the average total number of connections is $Nk_\mathrm{av}/2$, I have introduced a continuous parameter $P_0$, which is not greater than $ck_\mathrm{min}^{-\lambda}$, but is as close as possible to it.
Once the degree distribution is determined, actual networks are generated in the following way.
First of all, each node is allotted ``connecting arms'', the number of which is probabilistically chosen according to the above distribution.
Note that the total number of arms in the whole network is not fixed in this method, but its average will approach $Nk_\mathrm{av}$ as the number of networks in the ensemble becomes large enough.
Now, a random node is chosen as a seed of a cluster.
Then we pick another random node unconnected to the cluster and join one of its arms to a randomly chosen unconnected arm in the cluster.
This process is repeated until all nodes are connected to form one single cluster.
Finally, all unconnected arms are randomly paired, without doubly connecting any two nodes.\cite{even}

The critical transverse field $\Delta_c$ may be obtained using the finite-size scaling analysis on the fourth-order Binder cumulant\cite{binder1981a,hong2006a}
\begin{equation}
  U = 1 - \left[\frac{\left< m^4\right>}{3\left< m^2\right>^2}\right],
\end{equation}
where $m$ is the magnetization per spin, and $\left<\cdots\right>$ and $\left[\cdots\right]$ denote the thermal and network average, respectively.
In the vicinity of the quantum critical point, this quantity obeys a finite-size scaling form
\begin{equation}
  U = \tilde{U}\left( (\Delta-\Delta_c)N^{1/\nu'},TN^{z'} \right).
  \label{eq:scaling-binder}
\end{equation}
Since the size of a scale-free network is characterized by the total number of nodes $N$, instead of a length, I will denote our critical exponents with primed letters to distinguish them from the usual exponents $\nu$ and $z$.
In a system with $d$ spatial dimensions, they are related by $\nu'=d\nu$ and $z'=z/d$.
Since the upper critical dimension of the quantum phase transition in our quantum model is three,\cite{d_u} the mean-field values of these exponents become $\nu'=3/2$ and $z'=1/3$.
If $\Delta=\Delta_c$, the first argument in Eq. (\ref{eq:scaling-binder}) is zero and $\tilde{U}$ becomes a simple single-parameter scaling function.
Since $\tilde{U}$ has a peak, I can identify $\Delta_c$ by demanding that the maximum of $\tilde{U}$ as a function of the second argument should not depend on the system size $N$.
Then with an appropriate choice of $z'$, the curves for all system sizes should collapse onto a single curve within the scaling regime.

\begin{figure*}
\resizebox{\textwidth}{!}{%
  \includegraphics{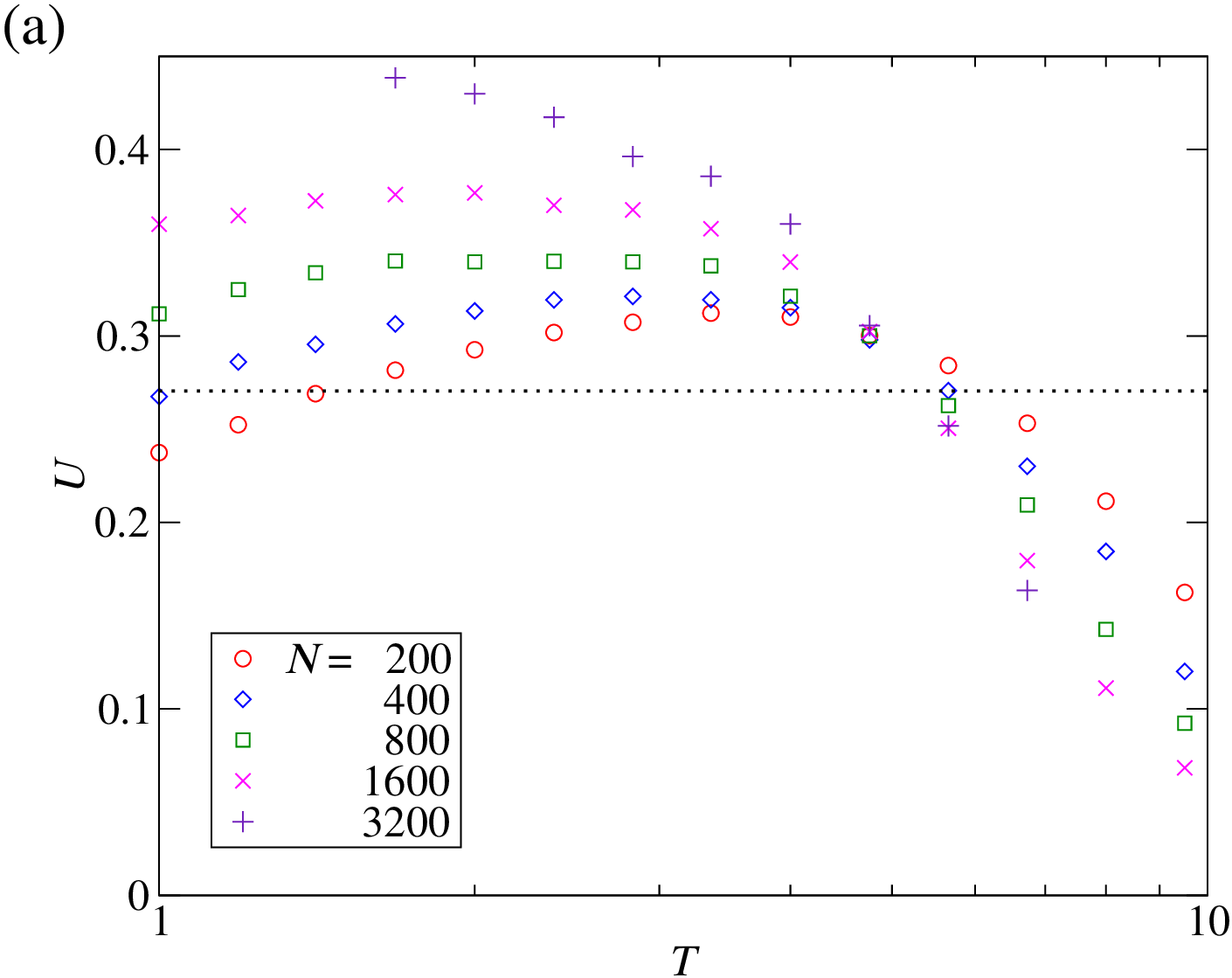}
  \includegraphics{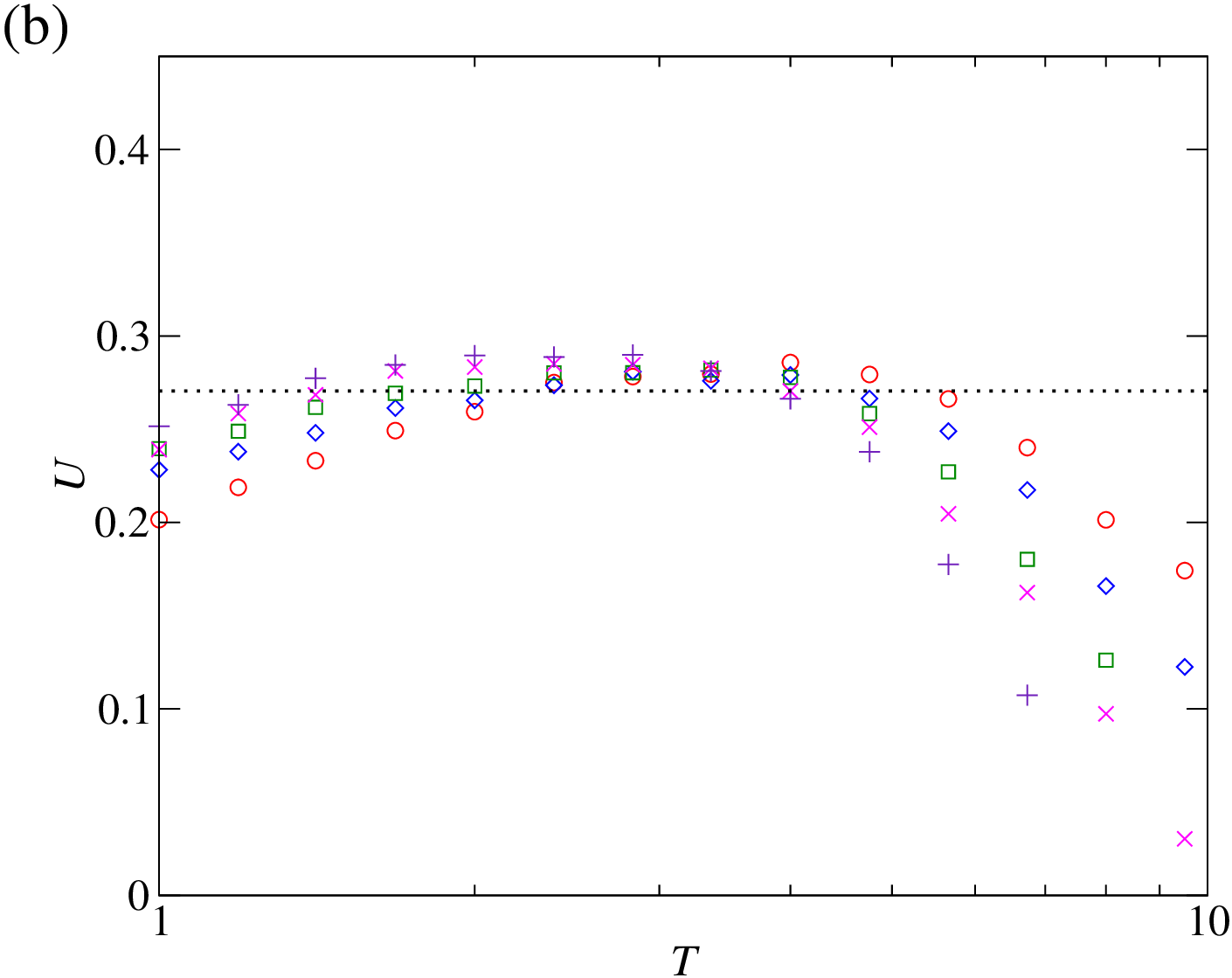}
  \includegraphics{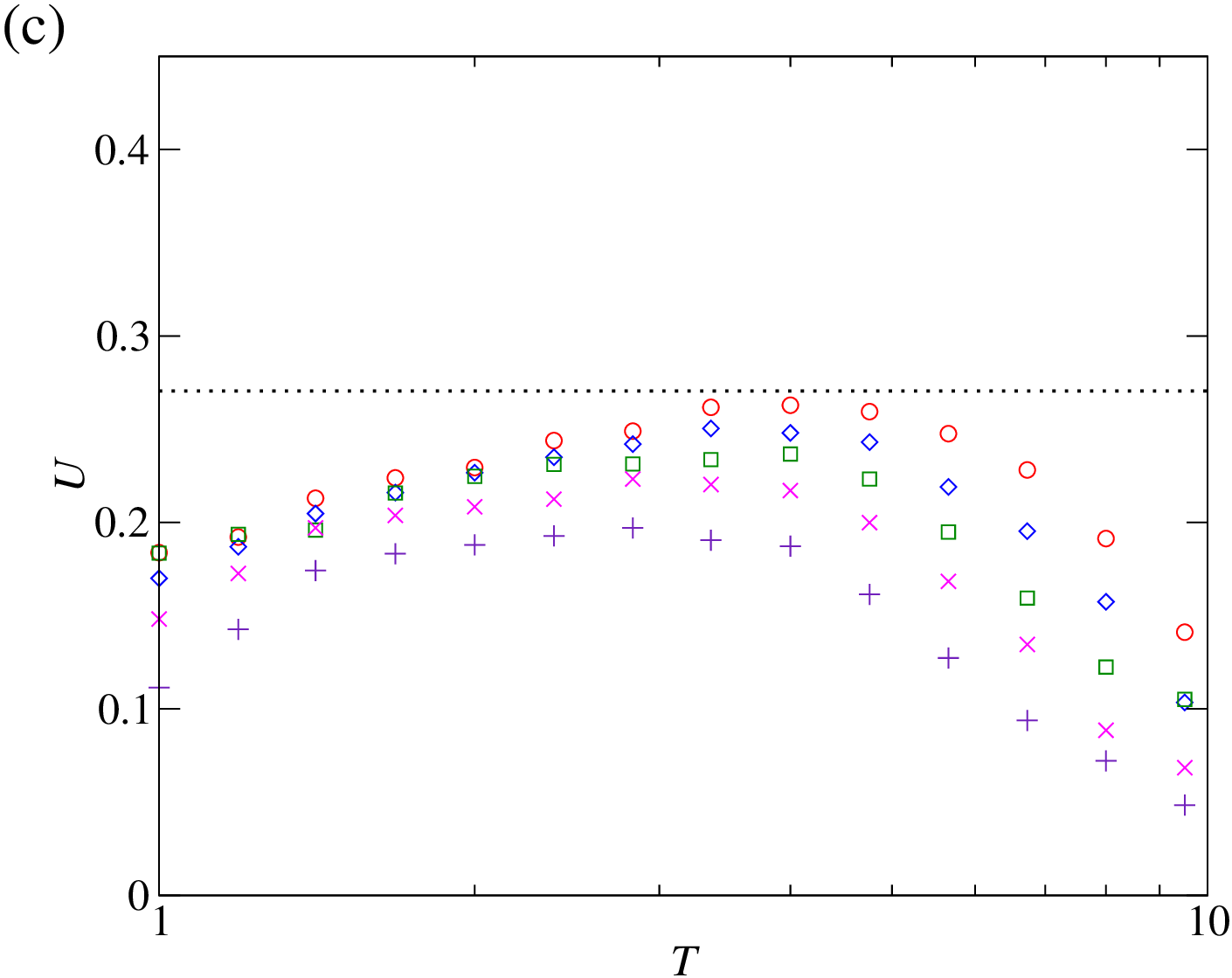}
}
\caption{
  (Color online)
  The Binder cumulant $U$ as a function of $T$ at (a) $\Delta=14.2$, (b) 14.35, and (c) 14.5 for $\lambda=6$ and $k_\mathrm{av}=7$. Different symbols are used to represent different system sizes as denoted in the plot. The dotted lines show the position of the mean-field value of the universal maximum $U^*\approx 0.270521$. The errorbars are omitted because they are negligible.
}
\label{fig:binder6}
\end{figure*}

\begin{figure*}
\resizebox{\textwidth}{!}{%
  \includegraphics{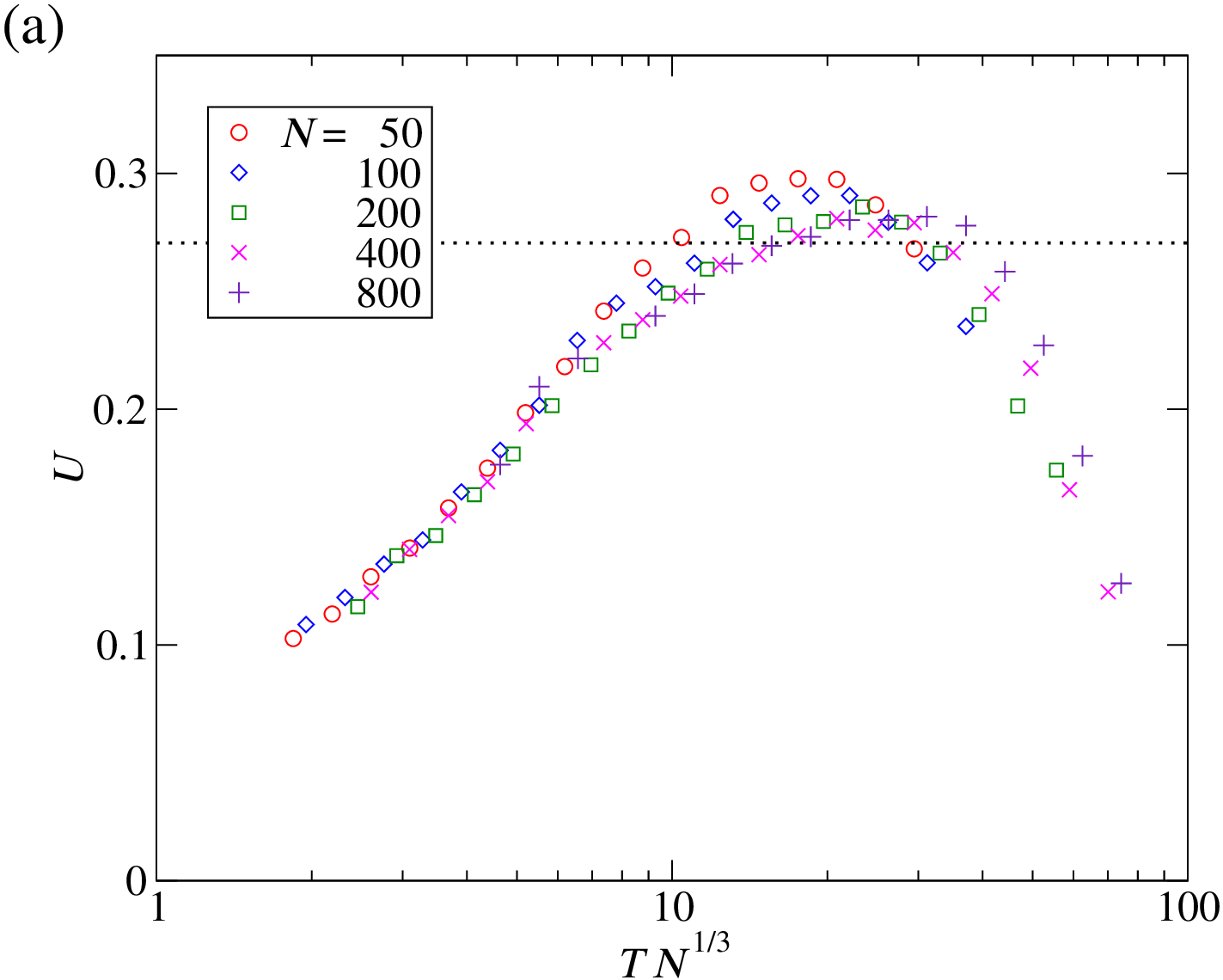}
  \includegraphics{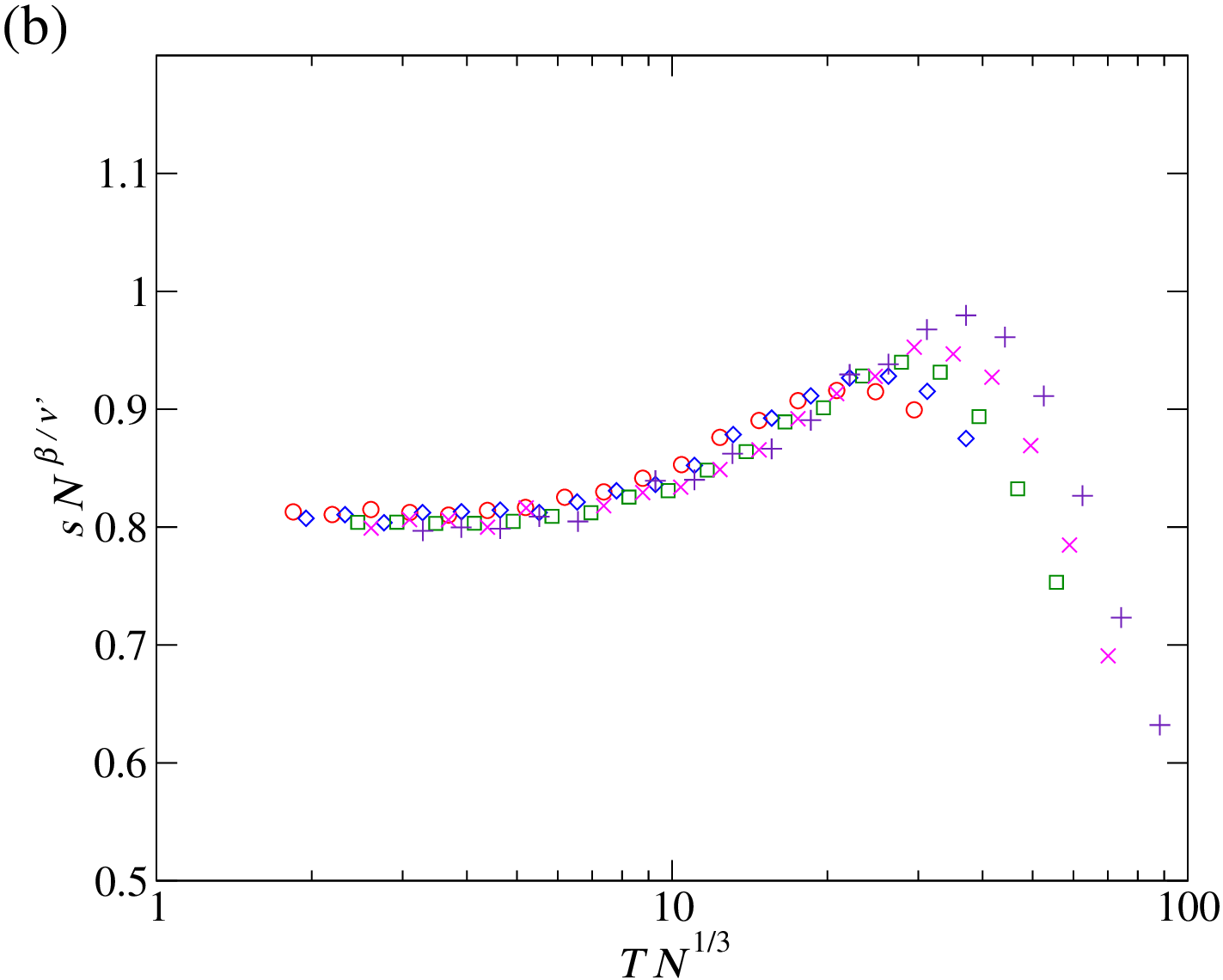}
  \includegraphics{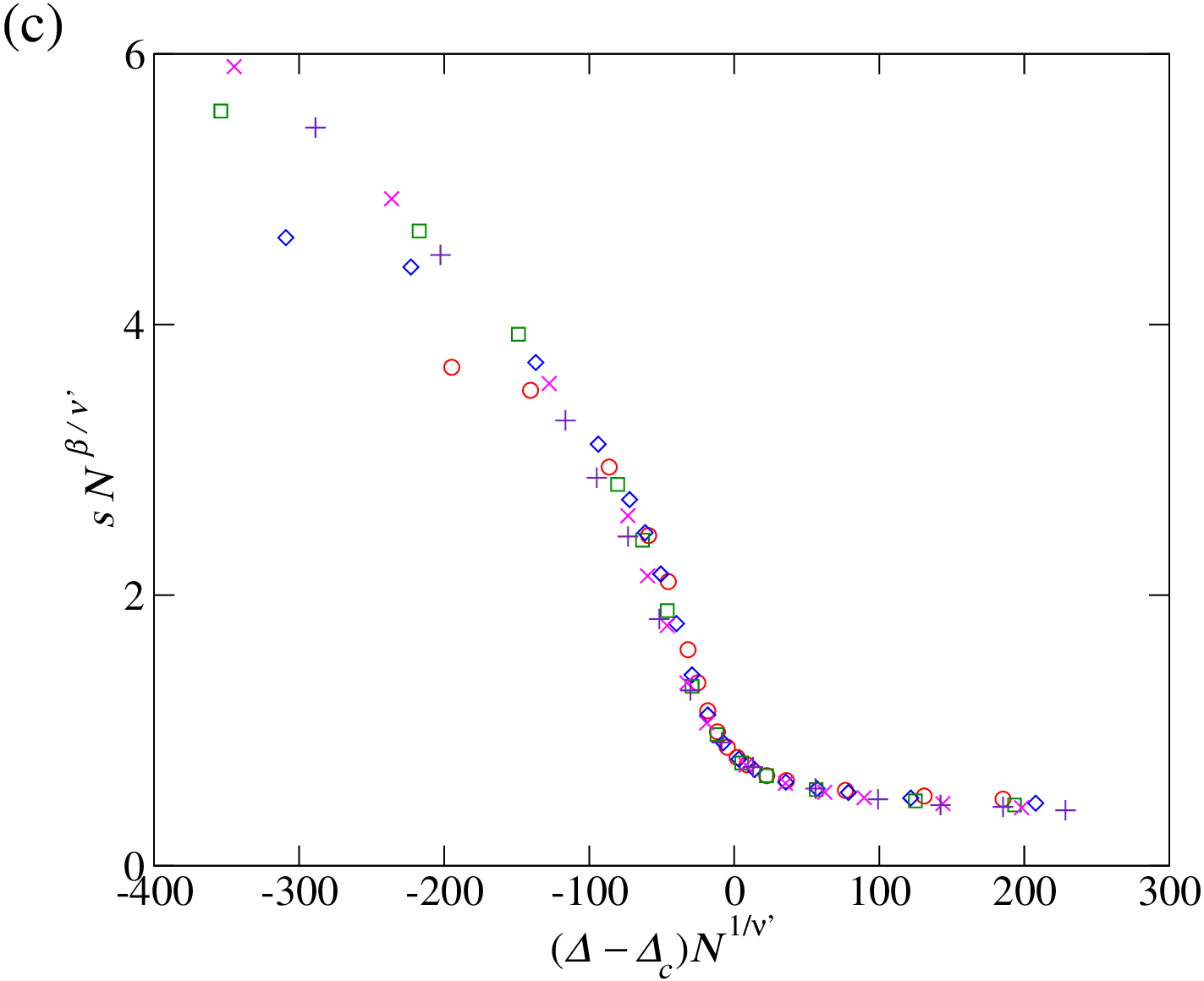}
}
\caption{
  (Color online)
  Data plots for $\lambda=6$ and $k_\mathrm{av}=7$: (a) $U$ and (b) $\tilde{s}$ vs $TN^{1/3}$ at $\Delta=14.35$. (c) $\tilde{s}$ vs $(\Delta-\Delta_c)N^{1/\nu'}$ at $TN^{1/3}=5.848035$. I assumed $\Delta_c=14.35$ and used the mean-field critical exponents $\nu'=3/2$ and $\beta=1/2$. The dotted line represents the mean-field universal maximum $U^*\approx 0.270521$.
}
\label{fig:scaling6}
\end{figure*}

\section{Results and analysis}
\label{sec:results}

\subsection{$\lambda>5$}

Let us first investigate the case where $\lambda=6$, for which the quantum critical point is expected to be in the mean-field universality class.
The results of the Binder cumulant calculations are shown in Fig. \ref{fig:binder6}.
Since the middle plot shows the smallest dependence of the maximum of $U$ on the system size $N$, we conclude that the critical transverse field is given by $\Delta_c=14.35\pm0.15$.
Note that the maximum of $U$ is a little greater than the mean-field universal value $U^*\approx 0.270521$,\cite{luijten1995a} but fairly close to it.
Figure \ref{fig:scaling6}(a) shows the Binder cumulant $U$ calculated at $\Delta_c$ as a function of $TN^{z'}$, using the mean-field dynamic critical exponent $z'=1/3$.
There are two things that call for special attention.
First, the data for different system sizes collapse onto a single curve for small $T$, but they start to fall apart near $TN^{1/3}\sim 7$.
This is a usual artifact of the finite-size effect.\cite{baek2011a}
Note that this may also account for the fact that our maximum overshot the universal value, albeit only by a little.
Second, it is not easy to accurately estimate $z'$ because we cannot use the data with $TN^{1/3}\gtrsim 7$, especially those near the peak.
Simply using the data in the low temperature scaling regime does not provide us enough information to determine $z'$ decisively.
Performing simulations with bigger systems may solve this problem, but it turns out to be very strenuous and time consuming with the current simulation method.

One way to overcome this difficulty and continue our analysis is to use a quantity that is not sensitive to the value of $z'$.
Below, I will use a method that has been suggested Baek {\it et al.} in Ref.~\onlinecite{baek2011a}.
Instead of the magnetization $m$, they have used the instantaneous magnetization per spin
\begin{equation}
  s = \frac{1}{N} \left[\left< \left| \sum_i \sigma_i^z(t) \right| \right>\right],
  \label{eq:s}
\end{equation}
taken at a given time $t$.
Due to the homogeneity in time, this quantity is actually independent of $t$.
The advantage of using $s$ lies in the fact that it saturates to a finite expectation value of the ground state as $T\rightarrow 0$.
As a consequence, it becomes independent of $T$ at low temperatures.
Therefore, it is expected to follow a simple single-parameter scaling form
\begin{equation}
  s = N^{-\beta/\nu'}\tilde{s}\left( (\Delta-\Delta_c)N^{1/\nu'} \right),
  \label{eq:scaling-s}
\end{equation}
which does not require prior knowledge of $z'$.
One can clearly see in Fig. \ref{fig:scaling6}(b) that indeed $\tilde{s}$ becomes independent of $T$ at low temperatures.
Picking a point in the flat region and sweeping $\Delta$ in the vicinity of $\Delta_c$, one may test the scaling property described in Eq. (\ref{eq:scaling-s}).
Below, I will use $TN^{1/3}=5.848035$, which falls inside the flat region of the plot in Fig. \ref{fig:scaling6}(b).

Figure \ref{fig:scaling6}(c) shows a plot of the scaling function $\tilde{s}\left((\Delta-\Delta_c)N^{1/\nu'}\right)$ where $TN^{1/3}$ is kept constant.
I have used $\Delta_c=14.35$ and the the mean-field critical exponents $\beta=1/2$ and $\nu'=3/2$.
It appears that all data for different system sizes collapse nicely onto a single curve in the vicinity of the critical point, and it supports the previous conjecture that the quantum critical point for $\lambda=6$ indeed belongs to the mean-field universality class.
From further analysis, I find that the same conclusion applies to all values of $\lambda$ as far as $\lambda>5$.

\begin{figure*}
\resizebox{0.75\textwidth}{!}{%
  \includegraphics{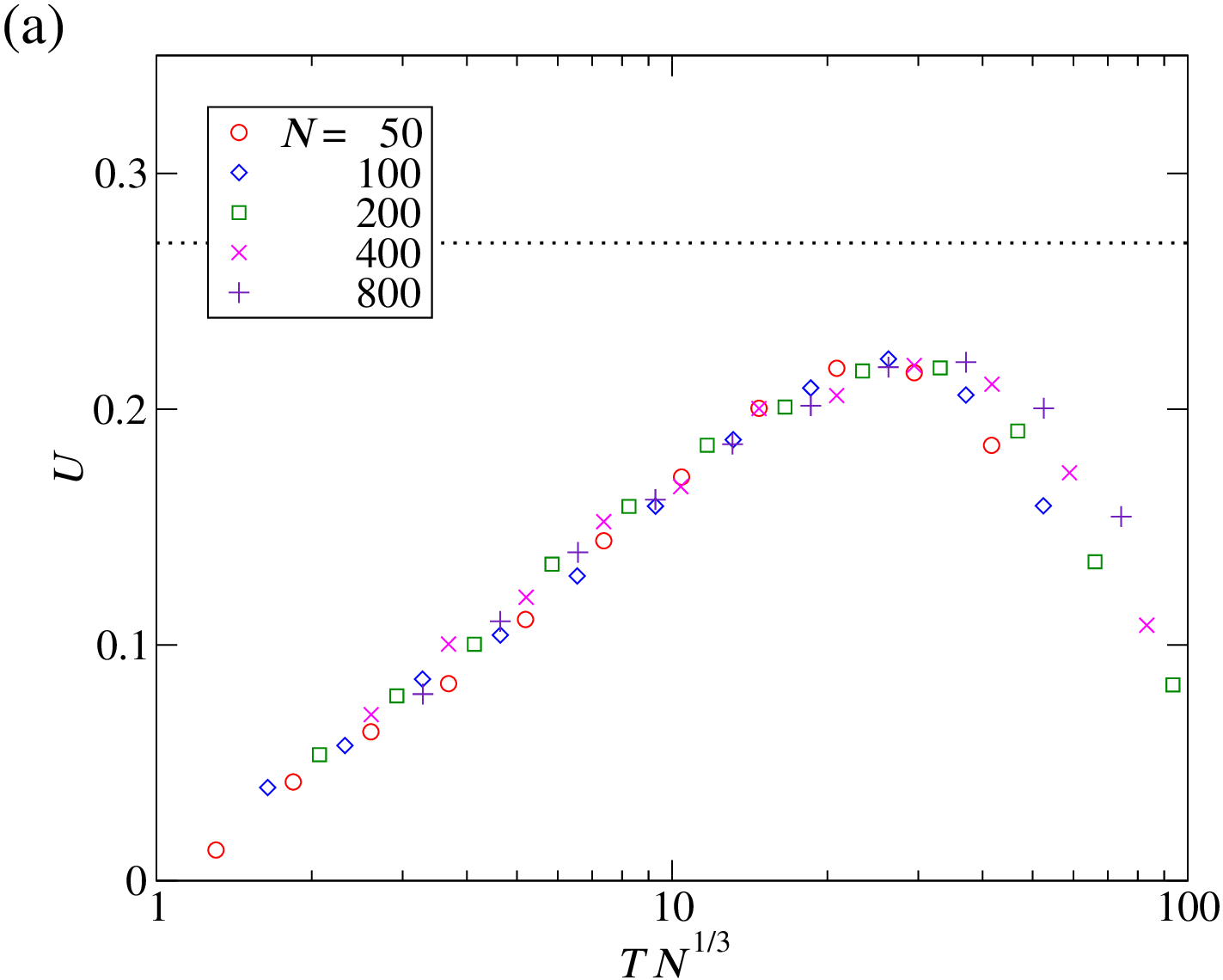}
  \hspace{0.35\textwidth}
  \includegraphics{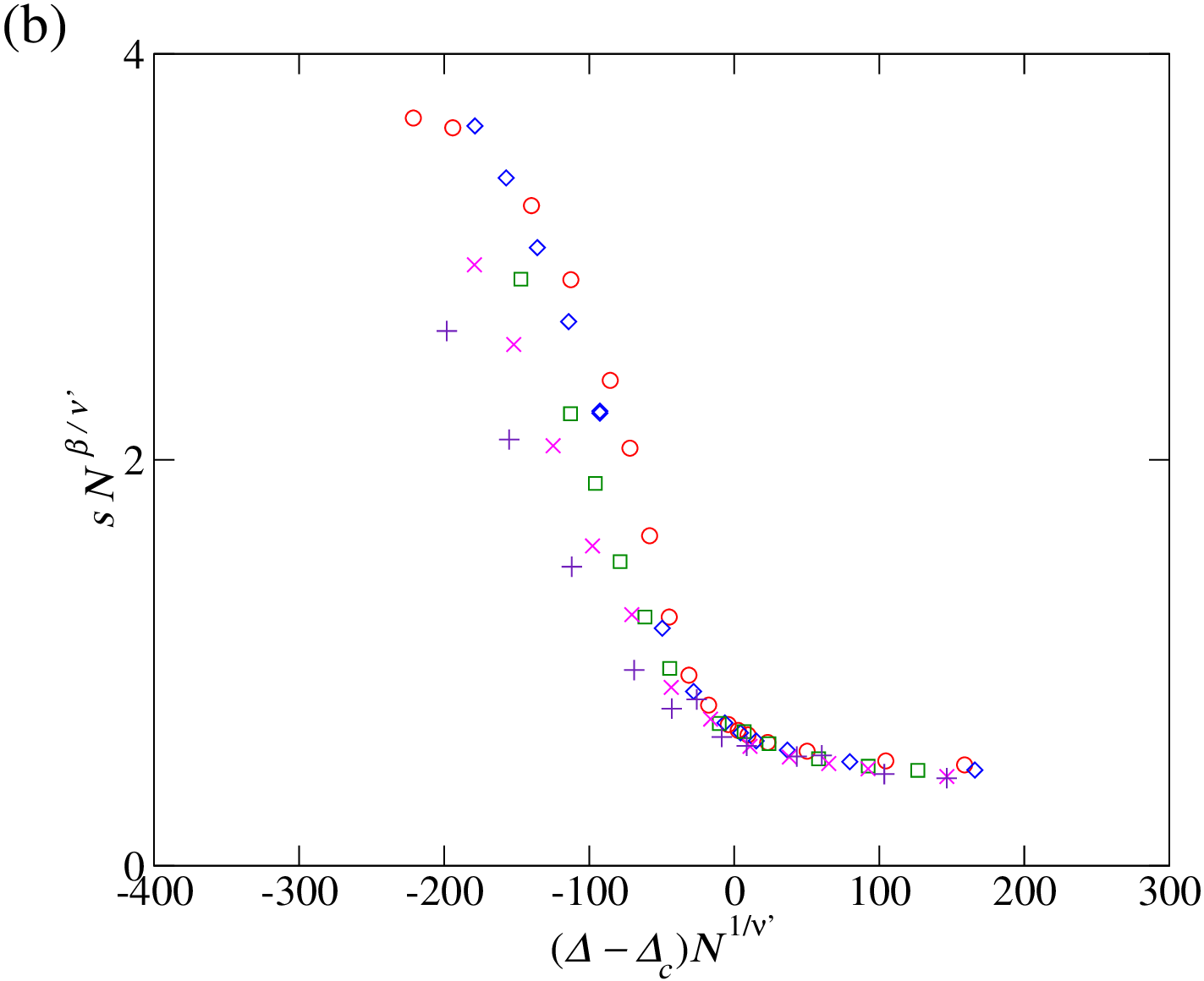}
}
\caption{
  (Color online)
  Data plots for $\lambda=4.5$ and $k_\mathrm{av}=7$: (a) $U$ vs $TN^{1/3}$ at $\Delta=16.3$; (b) $\tilde{s}$ vs $(\Delta-\Delta_c)N^{1/\nu'}$ at $TN^{1/3}=5.848035$. I assumed $\Delta_c=16.3$ and used the mean-field critical exponents. The dotted line represents the mean-field universal maximum.
}
\label{fig:scaling4.5}
\end{figure*}

\begin{figure*}
\resizebox{0.75\textwidth}{!}{%
  \includegraphics{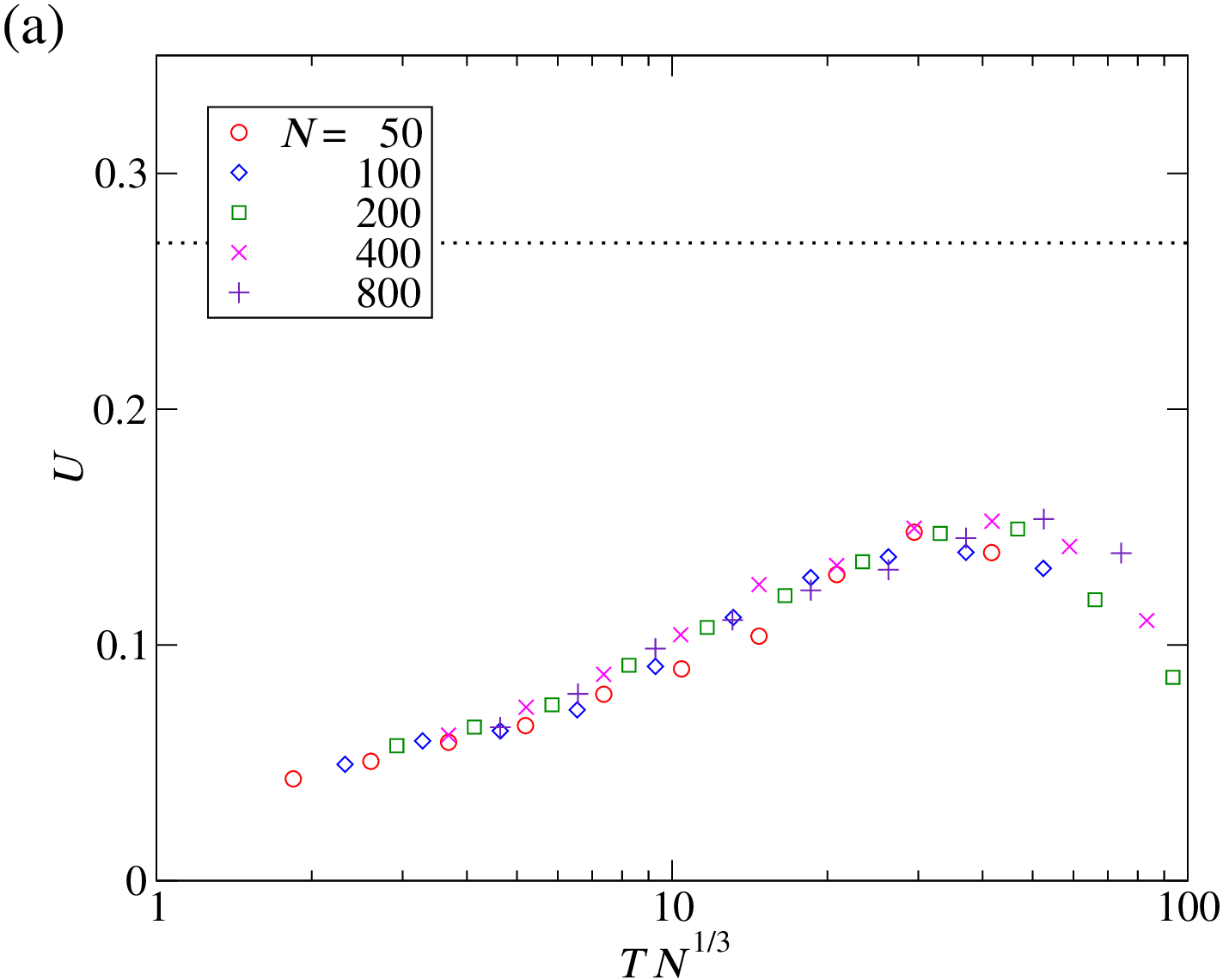}
  \hspace{0.35\textwidth}
  \includegraphics{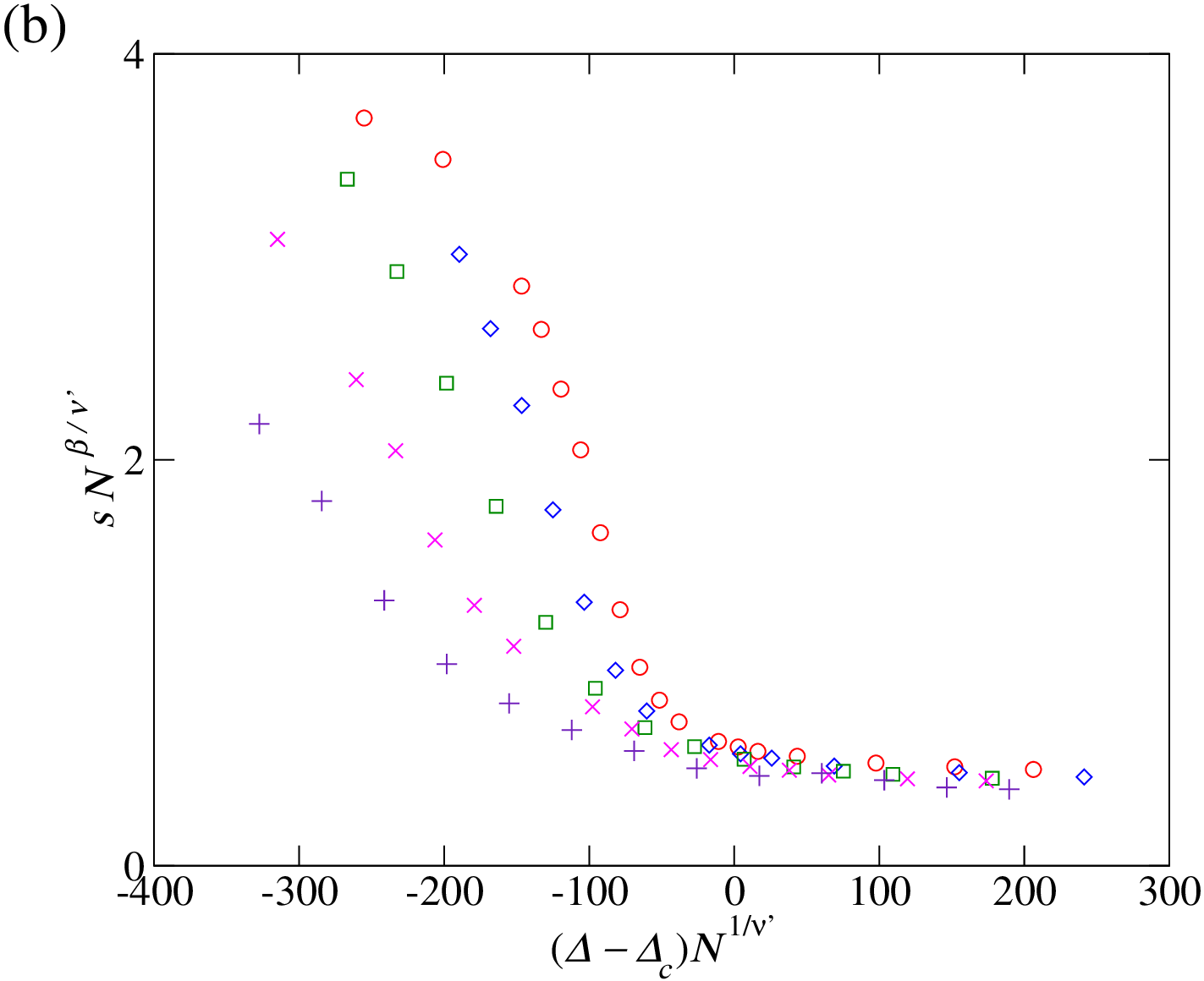}
}
\caption{
  (Color online)
  Data plots for $\lambda=4$ and $k_\mathrm{av}=7$. (a) $U$ vs $TN^{1/3}$ at $\Delta=18.8$. (b) $\tilde{s}$ vs $(\Delta-\Delta_c)N^{1/\nu'}$ at $TN^{1/3}=5.848035$. I assumed $\Delta_c=18.8$ and used the mean-field critical exponents. The dotted line represents the mean-field universal maximum.
}
\label{fig:scaling4}
\end{figure*}

\subsection{$3<\lambda<5$}

If $3<\lambda<5$, the {\em classical} critical point is not in the mean-field universality class, and the critical exponents are given by\cite{dorogovtsev2002a}
\begin{equation}
 \alpha = \frac{\lambda-5}{\lambda-3}, \quad \beta = \frac{1}{\lambda-3}, \quad \gamma = 1, \quad \nu' = \frac{\lambda-1}{\lambda-3}.
\end{equation}
However, the value of these exponents for the quantum critical point are yet to be discovered.
One interesting possibility is that the addition of the temporal dimension to the classical critical point might drive the quantum critical point to the mean-field universality class.
In order to test it, a similar analysis as in the previous subsection has been performed for $\lambda=4.5$.
The results are presented in Fig. \ref{fig:scaling4.5}.
First, we obtain the critical transverse field $\Delta_c=16.3\pm 0.2$ by carefully observing the maximum of $U$ as shown in Fig. \ref{fig:scaling4.5}(a).
This is greater than the value obtained for $\lambda=6$, which is a natural consequence of the fact that as $\lambda$ decreases, there are more hub nodes with very high degrees, which can enhance correlations even more.
The estimation of the dynamic critical exponent from $\tilde{U}$ is again inconclusive, and one cannot rule out the mean-field value $z'=1/3$. [Fig. \ref{fig:scaling4.5}(a)]
Notice, however, that the maximum of the Binder cumulant, which is estimated as $0.21\pm 0.01$, is conspicuously smaller than the mean-field universal value.
Just as I did for $\lambda=6$, I will assume the mean-field critical exponents and check its validity against the simulation data at a point in the flat region of $\tilde{s}$.
One can clearly see that scaling function $\tilde{s}$ plotted in Fig. \ref{fig:scaling4.5}(b) does not quite collapse into a single curve, when the mean-field critical exponents are used.

For $\lambda=4$, the discrepancy becomes even more obvious. The critical transverse field in this case is estimated as $\Delta_c=18.8\pm 0.2$.
As shown in Fig. \ref{fig:scaling4}(a), the Binder cumulant maximum is $0.15\pm 0.01$, which is almost only one half of the mean-field universal value.
One may also easily see that the scaling plot of $\tilde{s}$ using the mean-field critical exponents fails quite miserably. [Fig. \ref{fig:scaling4}(b)]
We thus come to the conclusion that the quantum critical point does not belong to the mean-field universality class for $\lambda=4.5$ and 4, and the deviation becomes more pronounced for the smaller value of $\lambda$.
Further analysis with other values of $\lambda$ in the domain $3<\lambda<5$ shows that the data near $\lambda\approx 5$ apparently agree with the mean-field theory, but the discrepancy becomes more and more pronounced as $\lambda$ gets smaller.
Due to the limit in the statistical error of the current study, however, it is not clear exactly when the model starts to deviate from the mean-field theory, or whether the change is abrupt or is just a smooth crossover.
The values of $\Delta_c$ and $U^*$ obtained from the current analyses are summarized in Table~\ref{table:summary}.

\begin{table*}[bth]
\centering
\begin{math}
\begin{array}{c||c|c|c|c|c|c|c|c|c|c}
\hline\hline
\makebox[3em]{$\lambda$} & \makebox[4em]{6.0} & \makebox[4em]{5.5} & \makebox[4em]{5.2} & \makebox[4em]{5.0} & \makebox[4em]{4.9} & \makebox[4em]{4.8} & \makebox[4em]{4.7} & \makebox[4em]{4.5} & \makebox[4em]{4.3} & \makebox[4em]{4.0} \\
\hline
\Delta_c & 14.35(15) & 14.6(2) & 14.8(2) & 15.0(2) & 15.2(2) & 15.4(2) & 15.7(2) & 16.3(2) & 17.2(2) & 18.8(2)\\
U^* & 0.28(1) & 0.28(1) & 0.28(1) & 0.28(1) & 0.27(1) & 0.26(1) & 0.24(1) & 0.21(1) & 0.18(1) & 0.15(1) \\
\hline\hline
\end{array}
\end{math}
\caption{The critical transverse field $\Delta_c$ and the universal maximum of the Binder cumulant $U^*$ for several values of $\lambda$. The numbers in parentheses denote the uncertainty in the last digits.}
\label{table:summary}
\end{table*}

\subsection{$\lambda<3$}

In the classical Ising model with $\lambda<3$, the critical temperature $T_c$ is infinitely large. Therefore, there is no ferromagnetic-paramagnetic phase transition, and the system is ordered at all temperatures. How the quantum model in this regime is different from the classical counterpart is an interesting question in its own right. Especially, the possibility of a quantum phase transition due to the transverse magnetic field is quite intriguing. Yet this subject is beyond the scope of the current research, and it will be discussed elsewhere.

\section{Summary and Discussions}
\label{sec:summary}

In this paper, I examined the quantum critical point of the transverse-field quantum Ising model on scale-free networks.
In order to check whether the quantum critical point belongs to the mean-field universality class, I assumed the mean-field values for the critical exponents and tested their validity.
Using the continuous-time quantum Monte Carlo simulation method, I calculated the Binder cumulant $U$ and the instantaneous magnetization $s$, both for $\lambda>5$ and $3<\lambda<5$.
From the observation of the maximum of $U$, the critical transverse field $\Delta_c$ was obtained.
While the estimation of the dynamic critical exponent $z'$ was problematic due to the finite-size effect, the use of the scaling behavior of $\tilde{s}$ allowed us to test the validity of the mean-field critical exponents.
For $\lambda=6$, I could confirm that the quantum critical point is in the mean-field universality class. For $\lambda=4.5$ and 4, however, the peak value of $U$ was substantially smaller than the mean-field universal maximum and the finite-size scaling behavior was clearly incompatible with the mean-field theory.
Further analysis showed that the data agree with the mean-field theory for all values of $\lambda$ as far as $\lambda>5$, but they start to deviate near $\lambda\approx 5$, and the discrepancy becomes more conspicuous as $\lambda$ decreases.

\acknowledgments
This research was supported by the Basic Science Research Program through the National Research Foundation of Korea funded by the Ministry of Science, ICT and Future Planning (NRF-2012R1A1A2041460).


\begin{thebibliography}{38}
\bibitem [{eve()}]{even}%
  \BibitemOpen
  \href@noop {} {}\bibinfo {note} {Because every connection involves two arms,
  the sum of all arms must be an even integer.}\BibitemShut {Stop}%
\end{thebibliography}
\end{document}